\documentclass[%
 twocolumn,
superscriptaddress,
showpacs,preprintnumbers,
prl,
]{revtex4-1}

\usepackage{graphicx}
\graphicspath{{pix/}}
\usepackage{dcolumn}
\usepackage{bm}
\usepackage[mathlines]{lineno}
\usepackage{amsmath}
\usepackage{mathtools}

\begin{document}

\title{Spin relaxation through Kondo scattering in Cu/Py lateral spin valves.}

\author{J. T. Batley}
\email{J.T.Batley@leeds.ac.uk}
\affiliation{School of Physics and Astronomy, University of Leeds, Leeds, LS2 9JT, UK.}
\author{M. C. Rosamond}
\affiliation{School of Electronic and Electrical Engineering, University of Leeds, Leeds, LS2 9JT, UK.}
\author{M. Ali}
\affiliation{School of Physics and Astronomy, University of Leeds, Leeds, LS2 9JT, UK.}
\author{E. H. Linfield}
\affiliation{School of Electronic and Electrical Engineering, University of Leeds, Leeds, LS2 9JT, UK.}
\author{G. Burnell}
\author{B. J. Hickey}
\email{B.J.Hickey@leeds.ac.uk}
\affiliation{School of Physics and Astronomy, University of Leeds, Leeds, LS2 9JT, UK.}
\date{\today}

\begin{abstract}
The temperature dependence of the spin diffusion length typically reflects the scattering mechanism responsible for spin relaxation. Within non-magnetic metals it is reasonable to expect the Elliot-Yafet mechanism to play a role and thus the temperature dependence of the spin diffusion length might be inversely proportional to resistivity. In lateral spin valves, measurements have found that at low temperatures the spin diffusion length unexpectedly decreases. By measuring the transport properties of lateral Py/Cu/Py spin valves, fabricated from Cu with magnetic impurities of $<$1 ppm and $\sim$4 ppm, we extract a spin diffusion length which shows this suppression below 30 K only in the presence of the Kondo effect. We have calculated the spin-relaxation rate and isolated the contribution from magnetic impurities. We find the spin-flip probability of a magnetic impurity to be 34\%. Our analysis demonstrates the dominant role of Kondo scattering in spin relaxation, even in low concentrations of order 1 ppm, and hence illustrates its importance to the reduction in spin diffusion length observed by ourselves and others.
\end{abstract}


\maketitle


Spin dependent phenomena underpin the field of spintronics, where an electron's spin is used in conjunction with its charge to create novel electronics and study spin transport. Progress within the field pushes the boundaries of solid state physics through applications of pure spin currents---a flow of spin angular momentum in the absence of a net flow of charge---presenting opportunities for information transport without dissipation from Joule heating. A thorough understanding of spin accumulation and transport effects is vital in real world exploitation of spintronics to provide new low power consumption and  non-volatile technologies\cite{Bader:2010gr,BehinAein:2010hj,Nikonov:2006kg}.

Nonlocal measurements in lateral spin valves (LSVs) are ideal for the study of spin transport due to the spatial separation of the spin and charge currents\cite{Johnson:1985jy,Jedema:2001ht}. The structure has been successfully used to measure spin transport through various metallic\cite{Ji:2004ik,Kimura:2004kc,Kimura:2007cq} and semiconducting\cite{Salis:2010ct,Suzuki:2011in} materials. LSVs consist of two ferromagnetic (FM) electrodes bridged by a nonmagnetic (NM) nanowire. The nonlocal resistance ($R_s$) is extracted for both parallel (P) and antiparallel (AP) configurations of the FM electrodes. The difference between P and AP is commonly labeled $\Delta R_s$ and is a measure of the accumulation and diffusion of spins within the NM. It is widely observed that the temperature dependence of $\Delta R_s$ within Cu/Py LSVs is non-monotonic, with a peak around 30 K\cite{Kimura:2008ce,Zou:2012gk,Villamor:2013ea,Erekhinsky:2012io,OBrien:2014cqa}. A similar behaviour is found in LSVs containing other materials\cite{OBrien:2014cqa,Idzuchi:2012cq,Mihajlovic:2010js}.

The non-monotonic behaviour of $\Delta R_s$ has prompted many theories focusing on either changes in the spin polarisation of the FM/NM interface $\alpha$\cite{OBrien:2014cqa} or scattering effects altering the spin diffusion length of the NM ($\lambda_{NM}$)\cite{Mihajlovic:2010js,Kimura:2008ce,Idzuchi:2012cq,OBrien:2014cqa,Zou:2012gk}. Arguments for the peaked behaviour of the nonlocal spin signal relating to transport in the NM require the addition of surface scattering from non-magnetic\cite{Mihajlovic:2010js,Kimura:2008ce,Idzuchi:2012cq} and magnetic sources\cite{Zou:2012gk,OBrien:2014cqa} which cause a reduction in $\lambda_{NM}$ below 50 K. The appearance of a thickness dependence of the temperature where the maximum spin signal occurs\cite{Kimura:2008ce} has been attributed to the temperature where the electronic mean free path is comparable to the thickness, increasing spin flipping events through surface scattering. However, experiments by Villamor \emph{et al.}\cite{Villamor:2013ea}, performed with two step lithography,  show no thickness dependence of the peak position in $\lambda_{NM}$ and as a result they introduced the possibility of magnetic impurities (MIs) as the origin of the low temperature reduction in $\lambda_{NM}$. 

The reduction of spin accumulation in LSVs fabricated through shadow deposition was shown by O'Brien \emph{et al.} to be a consequence of the diffusion of FM material into the NM at their interface\cite{OBrien:2014cqa}. Through measurements of LSVs fabricated with different FM/NM pairings, the reduction in $\Delta R_s$ is correlated to material combinations that exhibit a Kondo effect. The diffusion of FM material creates dispersed local moments, decreasing the injection/detection efficiency of the LSVs through Kondo scattering. This is suppressed in conventional two step lithography, where diffusion is minimised and there is no low temperature reduction of $\alpha$ but a qualitatively similar temperature dependence of $\Delta R_s$ is observed\cite{Villamor:2013hs}. Within these two step lithography samples the reduction in $\Delta R_s$ is singularly due to effects suppressing $\lambda_{NM}$ at low temperatures. A pivotal aspect here is the fabrication of the LSVs, and in particular, whether both FM and NM were deposited within the same vacuum cycle. These techniques alter the process by which magnetic impurities (MIs) enter the structure, either during the deposition\cite{OBrien:2014cqa,Zou:2012gk} or through the source material\cite{Villamor:2013ea}. In this Rapid Communication we show direct evidence linking the presence of the Kondo effect to the low temperature reduction in the spin diffusion length and provide a semi-quantitative explanation for this.

\begin{figure}[t]
\includegraphics[width=0.45\textwidth]{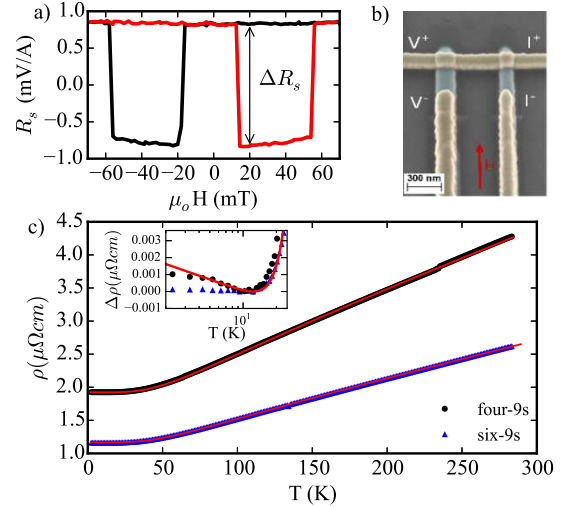}%
\caption{\label{device} a) The normalised nonlocal spin voltage measured at 5 K for a four-9s device with L = 425 nm. The red and black curves indicate increasing and decreasing field sweeps. b) An SEM image of a lateral spin valve showing the nonlocal measurement configuration and direction of applied field. The false colour indicates Py (blue) and Cu (orange). c) Resistivity data of the Cu spacer within a LSV. inset: $\Delta\rho = \rho(T) - \rho(T_{min})$ at low temperatures highlighting the Kondo minimum (black circles - four-9s Cu, blue triangles - six-9s Cu). The solid red line is a fit to equation (\ref{eq:whole}).}
\end{figure}

LSVs were fabricated on thermally oxidised SiO$_2$(100 nm)/Si substrates through a double dose electron-beam lithography and shadow deposition technique\cite{Rosamond:2015db}. Both materials are deposited in an UHV evaporation system at different angles to the substrate. The FM is deposited first (Py - Ni$_{80}$Fe$_{20}$) at 45$^{\circ}$, followed by the NM, copper (Cu) normal to the substrate. In order to investigate the the effect of MIs in spin relaxation, two sets of LSVs have been fabricated with different purity Cu. The two source materials were 99.99\% (four-9s) and 99.9999\% (six-9s) purity and the resistivity obtained from LSVs with different Cu is shown in Fig. \ref{device}(c). The lower quality four-9s Cu displays a higher resistivity due to the increase disorder from impurities. It is also evident from the low temperature data inset Fig. \ref{device}(c) that, down to the lowest temperature measured of 2 K,  only the four-9s samples have a resistivity minimum at $T_{min}$ = 11 K. A similar temperature dependence can be observed in systems with localization effects, however, here it is a consequence of MIs and Kondo scattering\footnote{See supplementary material for discussion on the origin of the minimum.}. 

Figure \ref{device}(b) is a scanning electron micrograph of a typical device. The FM electrodes have different widths along with a nucleation pad on one electrode to facilitate independent switching. The connecting copper wire of a four-9s and six-9s LSV has a cross-sectional area of $(1.36 \pm 0.05)\times 10^{-2} \mu m^2$ and $(0.82 \pm 0.05)\times 10^{-2} \mu m^2$ respectively \footnote{See supplementary material for further details on device dimensions.}. Devices have been fabricated with a wide range of separations between FM electrodes L, 400 nm to 1800 nm.

The nonlocal voltage is measured through a direct current injection method as a function of applied magnetic field. At each applied field value the nonlocal voltage-current (NLIV) characteristics are measured and the linear component is extracted to give $R_s$\footnote{Details on the NLIV and extraction of $R_s$ is given in the supplementary material}. At a large positive field, the two Py electrodes are aligned parallel (P) and produce a positive nonlocal signal as seen in Fig. \ref{device}(a). Through decreasing the magnitude of the applied field, at the coercivity of the easy electrode, there is a switch to a negative nonlocal signal as the magnetisation of the two Py electrodes align antiparallel (AP). By further increasing the magnitude of the applied field the P state is recovered when the coercive field of the second Py electrode is reached. The difference in $R_s$ between the P and AP states is indicated on Fig. \ref{device}(a) as $\Delta R_s$. This is the nonlocal spin signal and is proportional to the spin accumulation within the Cu. By applying the Valet-Fert \cite{Valet:1993dc} theory to a 1-D LSV geometry, an expression for the nonlocal spin signal is obtained:
\begin{equation}
\Delta R_s  = \frac{4  \alpha^2 R_F^2}{(1-\alpha^2)^2 R_N}\frac{e^{-\frac{L}{\lambda_N}}}{\big[1+\frac{2 R_F}{(1-\alpha^2)R_N}\big]^2-e^{-\frac{2 L}{\lambda_N}}}
\label{VF}
\end{equation}
The spin resistances for the FM and NM are defined as $R_F = \frac{\rho_F \lambda_F}{A_F},\ \ R_N = \frac{\rho_N \lambda_N}{A_N} $ where $\rho$ is the resistivity and A the cross-sectional area perpendicular to the spin current flow. $\lambda_N$, $\lambda_F$ and $\alpha$ are the spin diffusion lengths for the NM, FM and spin polarisation of the FM/NM interface respectively. 

\begin{figure}[ht]
\includegraphics[width=0.43\textwidth]{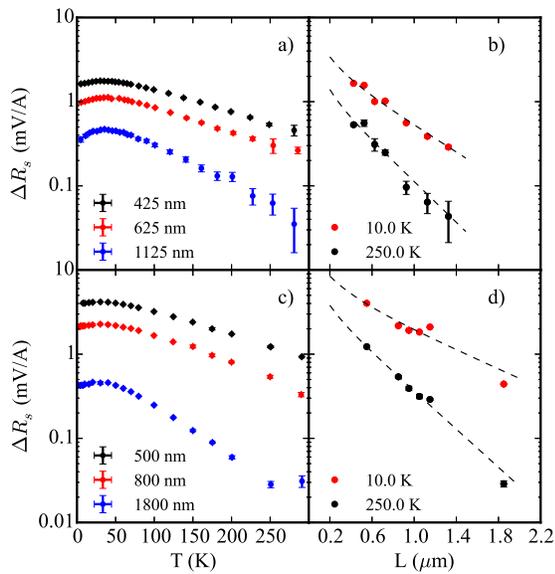}%
\caption{\label{DRvsT} Nonlocal spin signal in a) four-9s and c) six-9s LSVs of different L values as a function of temperature. A maximum is found at around 30 K. b) and d) Shows the decay of $\Delta R_s$ with L for four-9s and six-9s respectively. The dotted lines are fits to equation (\ref{VF}) }
\end{figure}

$\Delta R_s$ has been measured as a function of temperature within a helium flow cryostat, Fig. \ref{DRvsT}(a) and (c) shows the data for different magnetic electrode separations for the two different Cu purities. For all samples we observe a peak in the spin signal at around 30 K. In order to fit $\Delta R_s$ as a function of L, it is assumed that $\lambda_F \propto \sigma_F$ and fixed to a value of 5.5 nm at 4.2 K\cite{Steenwyk:1997dl} as similar to elsewhere\cite{Villamor:2013ea,Villamor:2013hs}. This leaves $\lambda_N$ and $\alpha$ as free fitting parameters. The fits are performed at different temperatures (see Fig. \ref{DRvsT}(b) and (d)) to obtain temperature dependent data on the spin relaxation in Cu. 


\begin{figure}[ht]
\includegraphics[width=0.48\textwidth]{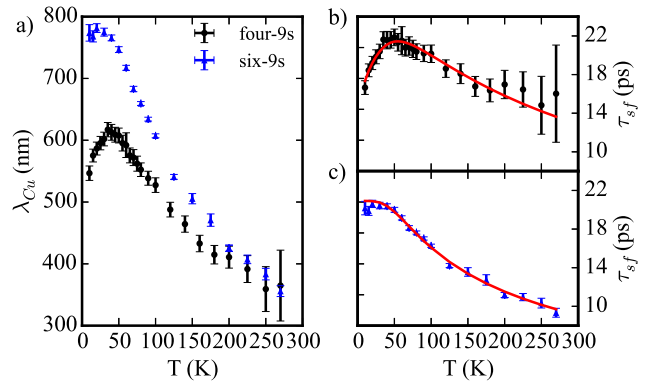}%
\caption{\label{sdl}a) Spin diffusion length (black circles - four-9s Cu, blue triangles - six-9s Cu) as a function of temperature. The values are obtained by fitting Eq. \ref{VF} to all electrode separations for each sample set. b) Spin flip scattering time for four-9s Cu as a function of temperature. Solid red line is a fit to equation (\ref{scattering}). c) Spin-flip scattering time for six-9s Cu as a function of temperature. The solid red line is a fit to equation (\ref{scattering}) assuming $(\tau_e^m)^{-1}=0$}
\end{figure}

Figure \ref{sdl}(a) shows the values for the spin diffusion length obtained from fits to equation (\ref{VF}) at various temperatures. Whilst the values for $\lambda_{Cu}$ lie well within the range of published values\cite{Jedema:2003cf,Casanova:2009ik,Zou:2012gk,Villamor:2013ea,Kimura:2008ce}, the two Cu purities show a significantly different temperature dependence. The six-9s purity Cu, in which no resistance minimum is observed, displays a monotonic increase and plateau at low temperatures within the experimental uncertainty. In contrast, the four-9s Cu sample set has a clearly visible peak around 35 K. This temperature dependence is extremely similar to other reports throughout the literature\cite{Kimura:2008ce,Villamor:2013ea,Mihajlovic:2010js}. 

The values obtained for the spin polarisation of the Cu/Py interface, between 0.4 and 0.3, are similar to those calculated for LSVs elsewhere\cite{Villamor:2013ea,Villamor:2013hs,OBrien:2014cqa}(not shown - see supplementary information). Both sets of LSVs have a reduction in $\alpha$ below 70 K due to the intermixing of Py and Cu at the interface\cite{OBrien:2014cqa}. 

Through sensitive resistivity measurements, and the presence of the Kondo effect\cite{Kondo:1964ip}, it is possible to detect the existence of MIs within a non-magnetic host. Analysis from the manufacturer shows the most significant impurities in the four-9s Cu are Ni and Fe with concentrations of 15 and 12 ppm respectively. Since the Kondo temperature for Ni impurities in Cu is greater than 1000 K, no resistance minimum is observed from Ni\cite{Inoue:1973ko,Schroeder:1965jp} and it is assumed that only Fe impurities contribute to this. With this we can use the empirical expression for Fe impurities within a Cu host $T_{min} = 115c^{1/5.3}$\cite{Knook:1962ta} to find the concentration of MIs to be c = 4 ppm This is lower than the source material due to a differential evaporation rate of the constituent elements.

The different contributions to momentum relaxation are obtained through fits to the resistivity data. A phenomenological model is built upon a combination of Bloch-Gr{\"u}neisen and Kondo terms to account for phonon ($\rho_p$) and magnetic impurity ($\rho_m$) scattering along with a constant term for nonmagnetic impurities ($\rho_i$):
\begin{subequations}
\label{eq:whole}
\begin{eqnarray}
\rho = \rho_i + \rho_p + \rho_m\label{subeq:2}
\end{eqnarray}
\begin{equation}
\rho = \rho_0 + K\Big(\frac{T}{\Theta_D}\Big)^5 \int_0^{\frac{\Theta_D}{T}} \frac{x^5\ dx}{(e^x-1)(1-e^{-x})} +\rho_K lnT\label{subeq:1}
\end{equation}
\end{subequations}
where $\rho_0$ is the residual resistivity due to non-magnetic impurities and defects, K is a material specific constant, $\Theta_D$ is the Debye temperature and $\rho_K$ is the Kondo resistivity. The fit of equation (\ref{eq:whole}) to the data is shown in Fig. \ref{device}(c) where it accurately reproduces all the features, including the Kondo minimum in the four-9s sample set. Values for $\Theta_D$ for high and low purity Cu are extremely close, $283.7 \pm 0.7$ K and $281.1 \pm 0.6$ K respectively. The constant K for the two sample sets are  $(6.13\pm0.02)\times10^{-7}$ $\Omega$m and $(9.79\pm0.03)\times10^{-7}$ $\Omega$m for six-9s and four-9s respectively. These values are dependent on the microstructure but are all in good agreement with those reported for Cu elsewhere\cite{Matula:1979vs,Poker:1982de}. The momentum relaxation rates are then each evaluated through the free electron Drude approximation $1/\tau_e = ne^2\rho_e/m_e$ where n is the conduction electron density and $\rho_e$ is the resistivity for the individual scattering contributions shown in equation (\ref{subeq:2}). 

The spin diffusion length can be defined as $\lambda_s = \sqrt{D \tau_{sf}}$ where D is the diffusion coefficient and $\tau_{sf}$ the spin-flip scattering rate. Using the Einstein relation $e^2ND=\sigma$ where N is the density of states at the Fermi energy and $\sigma$ is the electrical conductivity, we obtain D. Assuming a free electron model we take N = 1.8$\times 10^{28}$ states/eVm$^3$\cite{Ashcroft:1976ud} and the spin-flip scattering time is calculated as shown in Fig. \ref{sdl}(b) and (c). Within nonmagnetic metals the spin relaxation can be described through the Elliot-Yafet\cite{Elliott:1954eh,Yafet:1968bb} mechanism which, due to spin-orbit coupling, links the momentum and spin relaxation times $1/\tau_{sf} = \epsilon/\tau_{e}$, where $\epsilon$ is the probability of a spin-flip occurring at a momentum scattering event. From fits to the low temperature resistivity data we characterise the magnitude of the momentum relaxation rate due to Kondo scattering from MIs. This is present at all temperatures where the MIs maintain their moment, which can be up to room temperature\cite{Boyce:1976ft}. It has been shown that the spin-relaxation rate from MIs is proportional to their contribution to the momentum relaxation rate in a similar manner to the Elliot-Yafet mechanism\cite{Fert:1995ic}. Through applying Matthiessen's rule an expression for the total spin-flip scattering rate in terms of the individual momentum scattering rates and spin-flip probabilities is obtained as:
\begin{equation}
\frac{1}{\tau_{sf}} = \frac{\epsilon_{i}}{\tau_e^i} + \frac{\epsilon_{p}}{\tau_e^p} + \frac{\epsilon_{m}}{\tau_e^m}
\label{scattering}
\end{equation}
where the labels i, p and m represent non-magnetic impurities/defects, phonons and MIs. 

Samples fabricated with six-9s Cu do not show, within the experimental accuracy, a resistivity minimum and so $(\tau^m_e)^{-1}$ is assumed to be zero. A standard E-Y mechanism of spin relaxation through phonon and impurity scattering is used to fit $\tau_{sf}$ and reproduces the data well, shown in Fig. \ref{sdl}(c). Values for the spin-flip probability for phonon and nonmagnetic impurities of $\epsilon_p$ = (16.9$\pm$0.5)$\times 10^{-4}$ and $\epsilon_i$ = (17.2$\pm$0.1)$\times 10^{-4}$ respectively are in good agreement with those found in other LSV experiments \cite{Villamor:2013ea,Jedema:2003cf}. 

The lower purity four-9s $\tau_{sf}$ is not reproduced via only phonon and impurity scattering but requires the addition of Kondo spin relaxation which provides excellent agreement with the data, see Fig. \ref{sdl} (b). Values for the spin-flip probability for phonon and nonmagnetic impurities of $\epsilon_p$ = (8.1$\pm$0.5)$\times 10^{-4}$ and $\epsilon_i$ = (13.5$\pm$0.2)$\times 10^{-4}$ respectively are obtained. The spin-flip probability for MIs within Cu is obtained from the fit as $\epsilon_m = 0.34\pm0.03$, a value considerably larger than other contributors. An analytical expression for this can be obtained as\cite{Fert:1995ic}:
\begin{equation}
\epsilon_m = \frac{2}{3}\frac{S(S+1)\Big(\frac{J}{V}\Big)^2}{1+S(S+1)\Big(\frac{J}{V}\Big)^2}
\end{equation}
where S is the spin of the MI, V is the spin independent scattering potential and J the exchange coupling constant between the impurity and conduction electrons. The exchange coupling constant for Fe impurities in Cu has been previously been calculated as J$_{CuFe}$ = 0.91$\pm$0.2 eV \cite{Monod:1967kw}. From studies on CuMn dilute alloys a value of J/V = 0.133\cite{Fert:1981tt} is used with J$_{CuMn}$ = 0.4$\pm$0.1 eV\cite{Monod:1967kw}, and maintaining an equal impurity perturbation potential V, provides a ratio J/V = 0.30 for Fe impurities in Cu. Finally assuming for Fe$^{3+}$ impurities S=5/2, we obtain a value for $\epsilon_m$ of 0.29 in reasonable agreement with our result. Conventional spin-flip scattering in Cu is a consequence of the spin-orbit interaction, which can be parameterised through the coupling constant $\lambda \sim$ 0.2 eV\cite{Beuneu:1978hq,Fabian:1999uu}. However, Kondo scattering is a result of the s-d exchange interaction between the conduction electrons and the MI, which for Fe in Cu is J = 0.91 eV. The difference in the relative energy scales determines the strength of the spin flip scattering and accounts for the enhancement seen in Kondo scattering. It is this large spin-flip probability for MIs which makes them a dominant contribution to $\lambda_{Cu}$ even at temperatures above which their role is observed in the resistivity. LSVs fabricated with high purity Cu, with no observed Kondo effect, show no downturn within $\lambda_{Cu}$ at low temperatures, further supporting the role of Kondo scattering in spin relaxation within LSVs.    

The small impurity concentration of 4 ppm within the four-9s LSVs demonstrates a significant increase in spin relaxation at low temperatures. However this suppression of $\lambda_{Cu}$ is also measured in other studies where the Kondo minimum within the resistivity is not reported. To test the generality of our approach we repeat the analysis within \cite{Villamor:2013ea} with the addition of a magnetic spin relaxation term and using our value of $\epsilon_m = 0.34$. From this we are able to obtain a value for the Kondo resistivity $\rho_K$ leading to a MIs concentration of approximately 0.9 ppm through the relationship
\begin{equation}
\rho_K = c\Big(1+\frac{3zJ}{\varepsilon_F}\Big)\frac{3\pi m J^2 S(S+1)}{2ne^2 \hbar \varepsilon_F}
\end{equation}
where z is the number of conduction electrons per atom and $\varepsilon_F$ is the Fermi energy taken to be 1 and 7 eV respectively for Cu. This very small amount can clearly be seen as a significant contribution to the spin relaxation but is on the limit of where  a Kondo minimum within the resistivity would be observed, with a predicted $T_{min} \approx 8$ K and below the data reported in reference \cite{Villamor:2013ea}. 

From temperature dependent measurements of the spin current within LSVs of different purity Cu, the contributions to momentum relaxation from impurities, phonons and the Kondo effect are isolated. These have been related to the spin relaxation rates through an Elliot-Yaffet like model in combination with Matthiessen's rule. The observed values for spin-flip probabilities for phonon and non-magnetic impurities agree with those published for LSVs elsewhere. We show that even extremely small magnetic impurity concentrations can cause a significant reduction in the spin diffusion length at low temperatures due to the large probability of a spin-flip event with Kondo scattering. Our extension to include MIs accurately reproduces the spin relaxation in this low temperature regime providing a semi-quantitive method for analysis of the effect of MIs on pure spin currents and well describes a common feature seen throughout the literature. 
 
This work was funded by EPSRC and the Henry Ellison scholarship awarded by the University of Leeds. The data presented can be found at http://doi.org/10.5518/12.
\bibliography{bibliography}

\end{document}